# Energy dependent analysis of eta-meson photoproduction on protons and neutrons without background in the energy region up to 3.4 GeV

V.A. Tryasuchev

*Tomsk Polytechnic University, 634050 Tomsk, Russia*

tva@tpu.ru

Abstract

Observed for the processes $\gamma p \to \eta p$, $\gamma n \to \eta n$ are calculated throughout the resonance energy region using the single isobar model without a background. The connection between invariant photoexcitation amplitudes of the resonances on neutrons and protons are calculated. The results are discussed in terms of a comparison with experimental data and calculations performed by other authors.

New methods which are developed for detection of neutral mesons produced with tagged photons allow one to carry out systematic measurements of different observables for η-photoproduction both on protons [1–9]

$$\gamma + p \to \eta + p \qquad (1)$$

and on neutrons [10-12]

$$\gamma + n \to \eta + n. \qquad (2)$$

As a theory basis used to describe the reactions (1) and (2) one usually adopts the isobar models [13–19] which are extended, in some cases, to higher energies using the Regge model approach [18]. Our model contains only *s*-channel resonances. The corresponding resonance multipoles were taken in the Breit-Wigner form, but in contrast to other models which use the Walker parametrization [20] for vertices and propagators, in the present calculation we used the Mongtomery parametrisation [13, 21]. The nonresonant background which consists of the nucleon pole term and ρ- and ω-exchange in the *t*-channel [15, 16] was not included. This is because it leads to infinitely large cross section at high energies. One of the methods to avoid this problem is to introduce formfactors into the hadronic vertices ρ*NN* and ω*NN*, as is done for example in Refs. [17, 18]. However, if the η*NN* coupling contains a pseudoscalar part, then the validity of the formfactors in background terms is rather questionable. At the same time these terms will determine the role of resonances. It is therefore reasonable to adopt for the processes (1–2) a pure resonance model, as is done in the present work.

In its first version the model included 12 resonances [19], which were further reduced to only 8 ones [22, 23, 24, 25]. The resonances entering the latest version of our model are listed in Table 1. In the columns 4 and 5 of the table we give adjustable parameters $\gamma^E$ and $\gamma^M$, which are the product of hadronic and electromagnetic vertices of the corresponding resonance states [25]. They are related to the helicity amplitudes $A_\lambda$ ($\lambda$ = 1/2, 3/2) which in turn are related to the quantities $\xi_\lambda$:

$$\xi_\lambda = \sqrt{\frac{km\Gamma_{\eta p}}{qW_r\Gamma_r^2}} A_\lambda. \quad (3)$$

The latter were firstly introduced in Ref. [16] where it was shown that the values of $\xi_\lambda$ depend weakly on the chosen model. In Eq. (3), $m$ is the nucleon mass, $k$ and $q$ are absolute values of the photon and meson center-of-mass (c.m.) momenta, which correspond to the resonance position $W_r$ of the total c.m. energy, $\Gamma_{\eta p}$ and $\Gamma_r$ are respectively the partial and the total decay width of the resonance $r$. The absolute values $|\xi_\lambda|$ determine not only the amplitudes $A_\lambda$ of each individual resonance, but also its contribution to the resulting amplitude for $\gamma N \to \eta N$ (see Refs. [16, 25]).

**Table 1**. Resonance parameters obtained from the fit to the existing $\gamma p \to \eta p$ data. The preliminary data from Ref. [9] were not included into the fit.

| $N^*$-resonances | $W_r$, MeV | $\Gamma_r$, MeV | $\gamma^E$, MeV | $\gamma^M$, MeV | $\xi_{1/2}$, $10^{-1}$ GeV$^{-1}$ | $\xi_{3/2}$, $10^{-1}$ GeV$^{-1}$ |
|---|---|---|---|---|---|---|
| $S_{11}$(1535) | 1532 | 161 | 2.256 | — | 2.520 | — |
| $S_{11}$(1650) | 1642 | 140 | - 0.760 | — | - 0.975 | — |
| $S_{11}$(1830) | 1828 | 155 | 0.190 | — | 0.220 | — |
| $P_{11}$(1440) | 1440 | 350 | — | 0.250 | — | — |
| $P_{13}$(1720) | 1724 | 173 | - 0.080 | 0.560 | 0.267 | 0.665 |
| $D_{13}$(1520) | 1520 | 120 | 0.185 | 0.400 | - 0.064 | 0.119 |
| $D_{15}$(2150) | 2140 | 382 | 0.058 | 0.410 | 0.191 | - 0.143 |
| $F_{15}$(1680) | 1685 | 130 | 0.174 | 0.095 | 0.036 | 0.129 |
| $F_{17}$(1990) | 1999 | 312 | 0.025 | 0.330 | 0.139 | 0.154 |

The differential cross sections for η photoproduction on protons presented by GRAAL [4] and MAMI [8] are in general agreement with each other in the energy region from threshold up to the photon energy $K_0$ = 1500 MeV. In Ref. [24] the parameters of our model were adjusted to the GRAAL data [4]. The values which come out of this fit are only slightly different from those listed in Table 1. In the present work, to fit the low energy data ($K_0 <$ 1500 MeV) we chose the data [8], since they have better statistic. The corresponding results are presented in Fig. 1. The data from [4] and [8] are compared to each other in Fig. 1 on the panel, corresponding to the energy $K_0$ = 1425 MeV. Although one notes general agreement between the measurements, there is a small difference at the forward η angles. There is no need to present here all the existing data to show the quality of their agreement with our calculations (this was already done in some works, for instance in Refs. [27] and [28]). Therefore, we show our results and compare them with the most recent data with a definite step in energy.

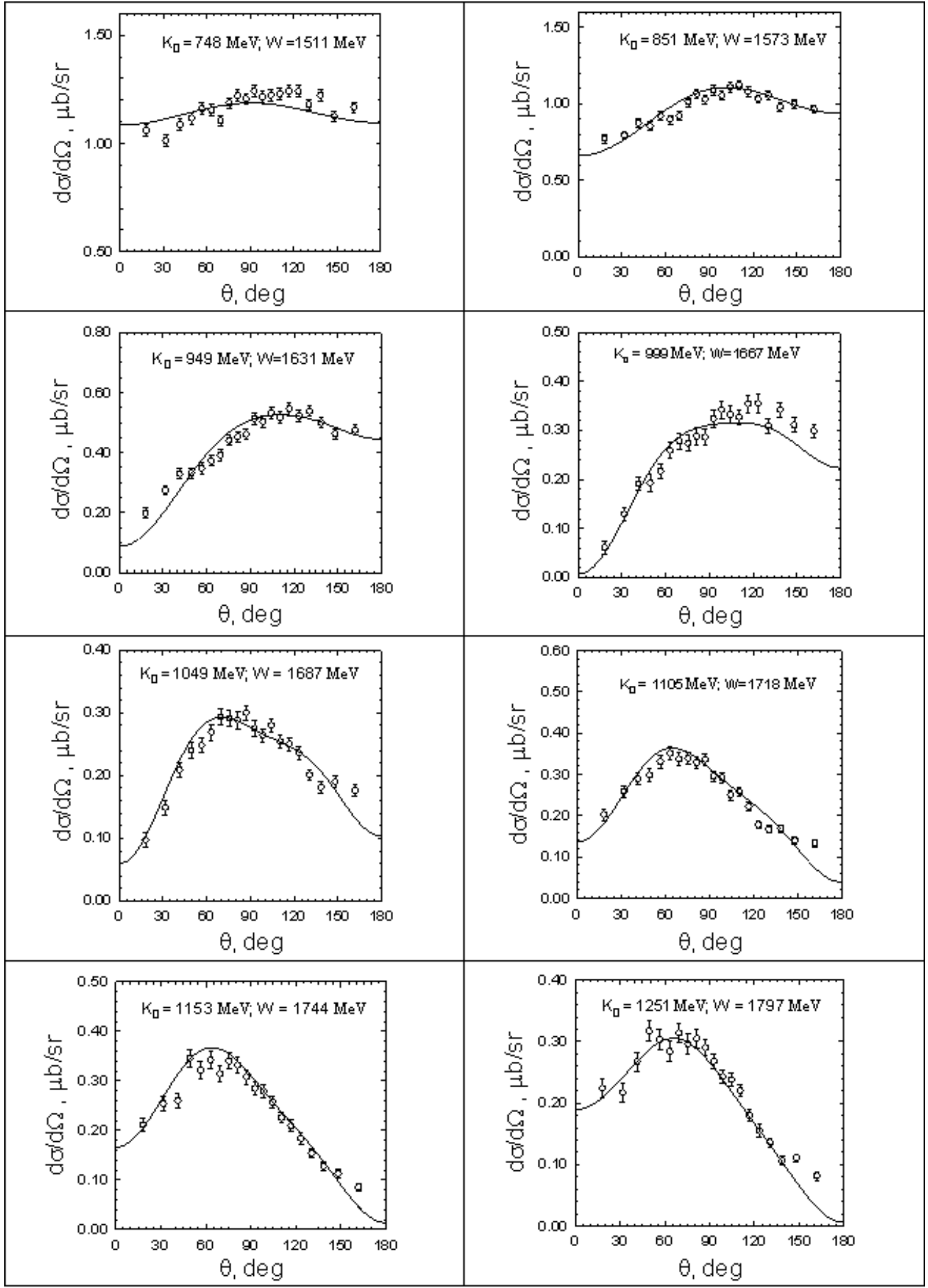

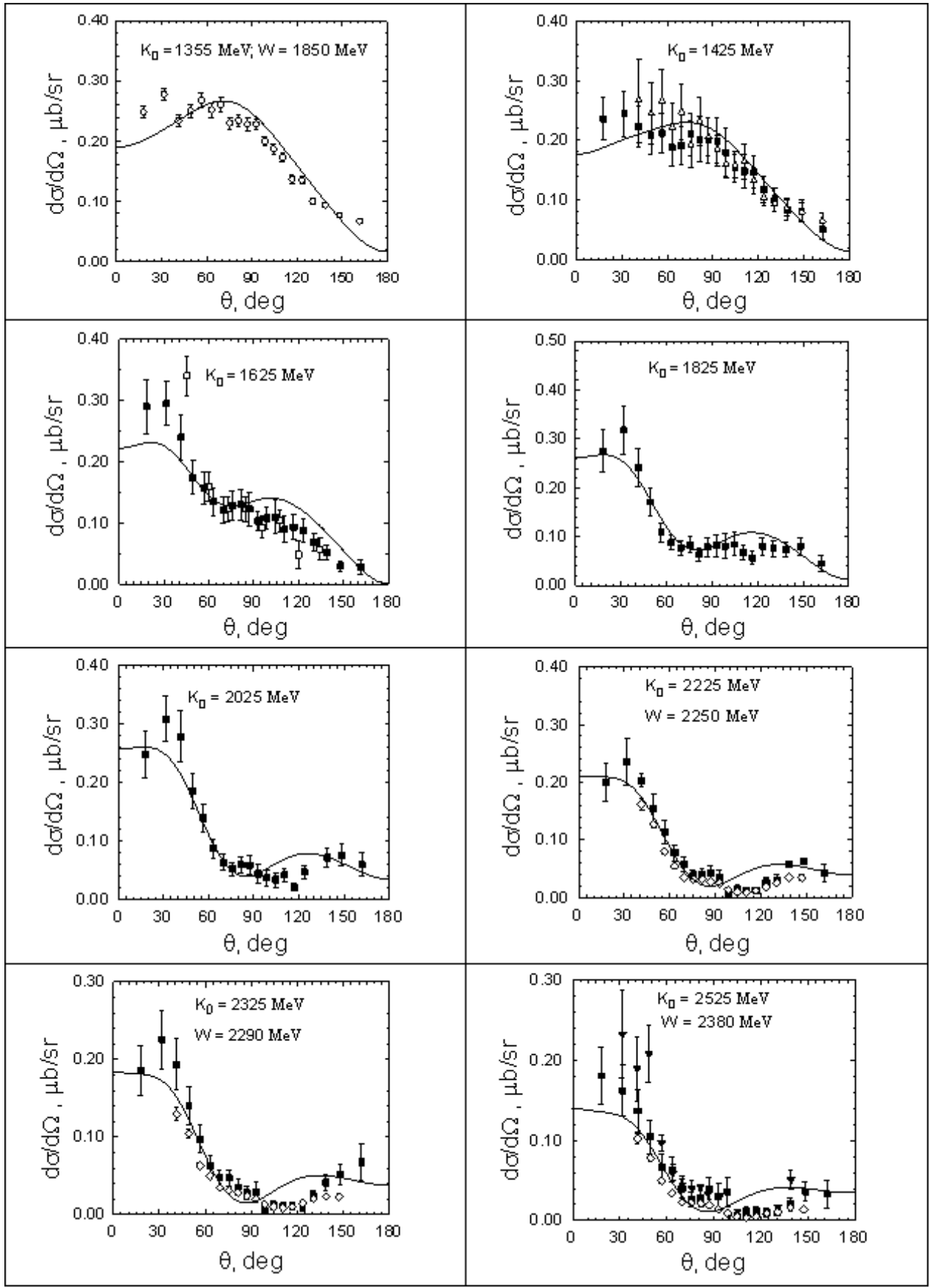

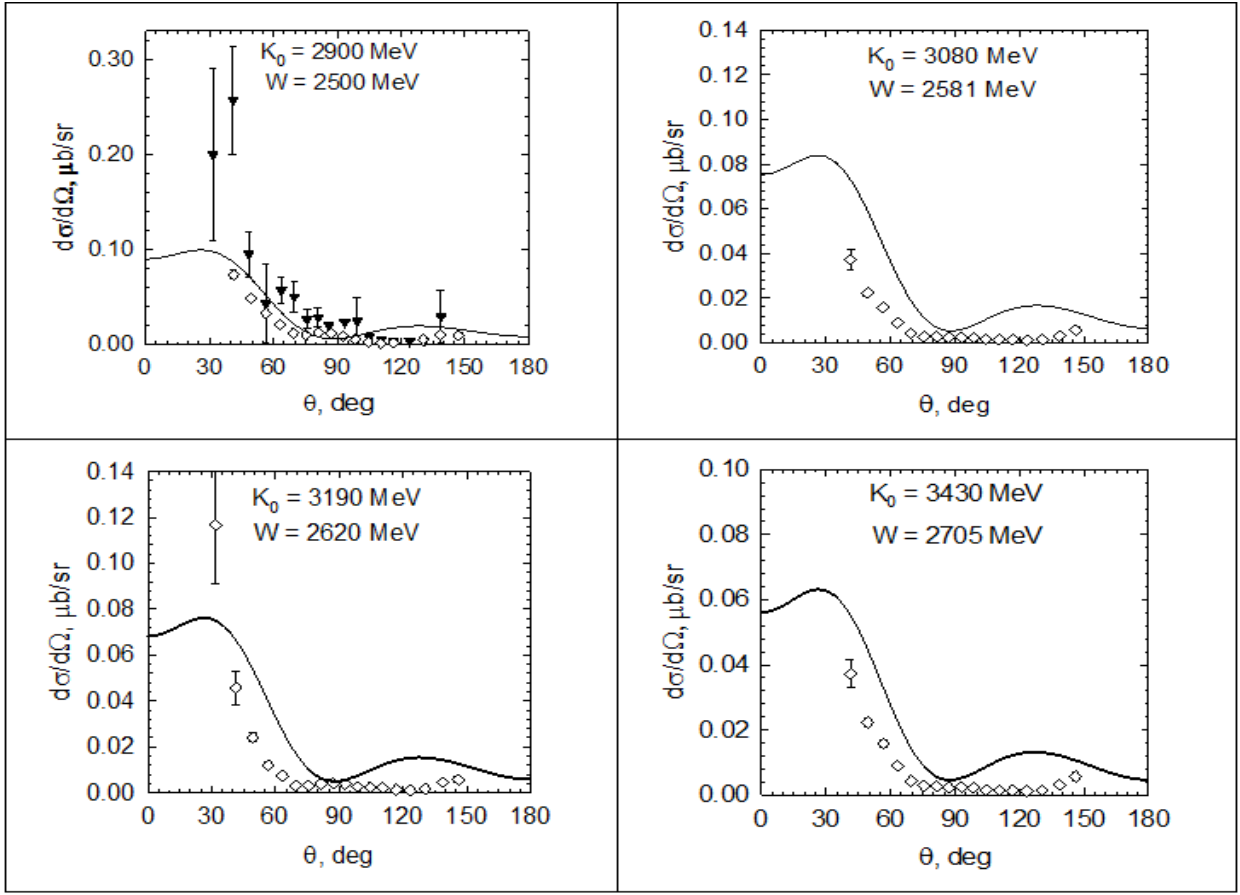

Fig. 1. Differential cross section for $\gamma p \to \eta p$ as function of the $\eta$- angle in the center-of-mass (c.m.) system for different photon energies $K_0$ ($W$ is the corresponding total c.m. energy). Experimental data are taken from Ref. [8] (○), [4] (Δ), [7] (■), [6] (◊), [1] (□), and [3] (▼)

At higher photon energies, $K_0 > 1500$ MeV, we additionally included the experimental differential cross sections from Refs. [6] and [7]. In the range $1500 < K_0 < 2225$ MeV only the data [7] were included into the fit. Among the three series of measurements presented in [7] which correspond to the three most important $\eta$ decay modes and respectively to the three different $\eta$ detecting methods, we chose the series in which the products of two $\eta$ decay modes were detected simultaneously (Fig. 11 in Ref. [7]). Although the measurements presented in [6] and [7] demonstrate visible differences, thus leading to some difficulties in their interpretation, these differences seem to be not crucial up to $K_0$=2325 MeV. Above this energy the differential cross sections of Ref. [6] are systematically below the corresponding values given in Ref. [7] (see Fig. 1). For $K_0 > 2900$ MeV, only the data from Ref. [6] are shown. One should note that reasonable description of the data at high energies was achieved only after the resonance $D_{15}(2150)$ had been introduced. This resonance does not enter the last Particle Data Group compilation [29], but it was mentioned in Ref. [28] as a possible nucleon excited state. On the last three panels the agreement between the data and our calculation is only qualitative. At the same time, one may expect relatively large systematic error in this region.

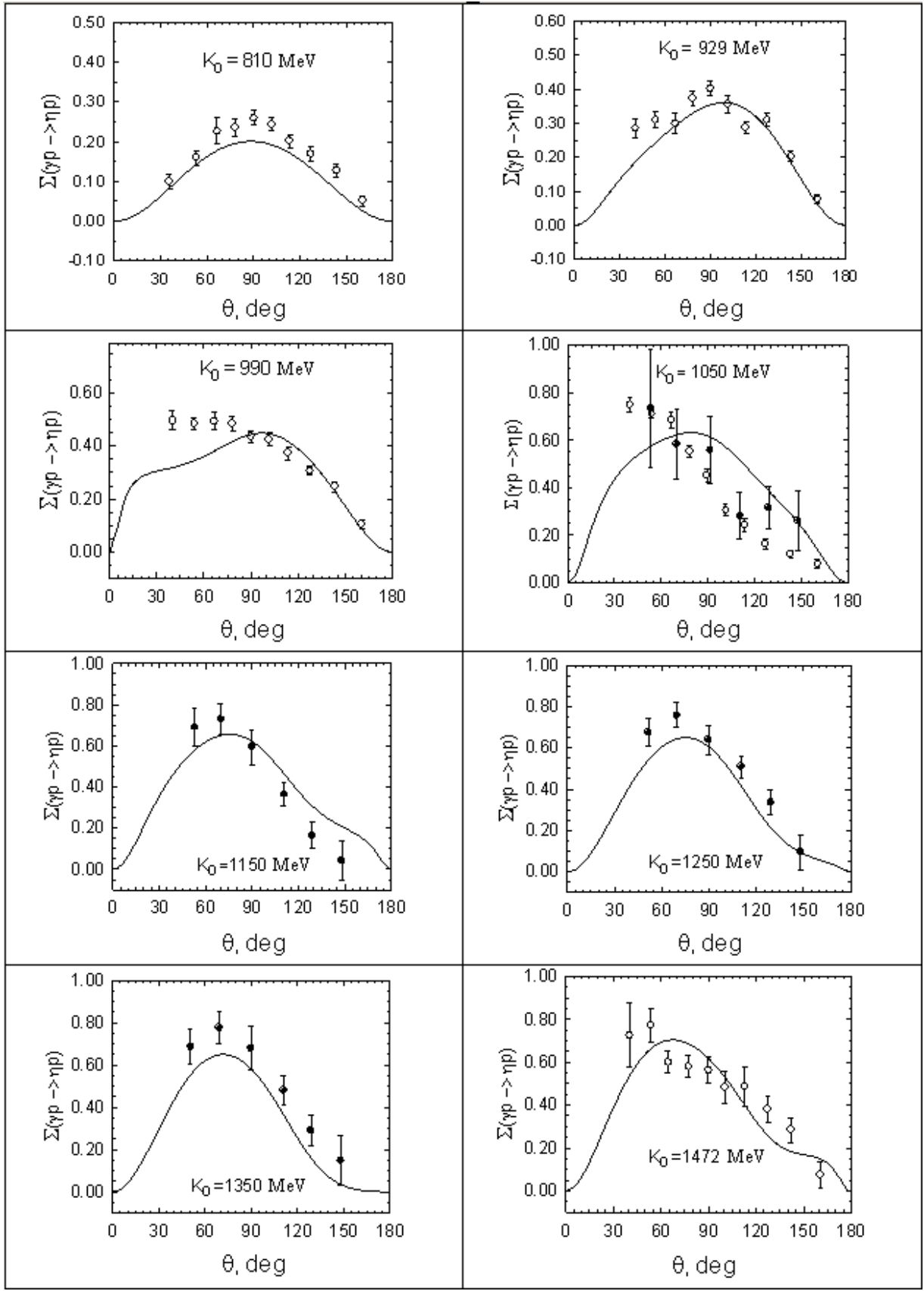

Fig. 2. $\sum$– asymmetry for the process $\gamma p \to \eta p$ as function of $\eta$ c.m. angle for different photon energies $K_0$. Experimental data are from Ref. [4] (○) and [5] (●)

The calculated beam asymmetry for $\gamma p \to \eta p$ is presented in Fig. 2, where it is also compared with the experimental results from Refs. [4] and [5].

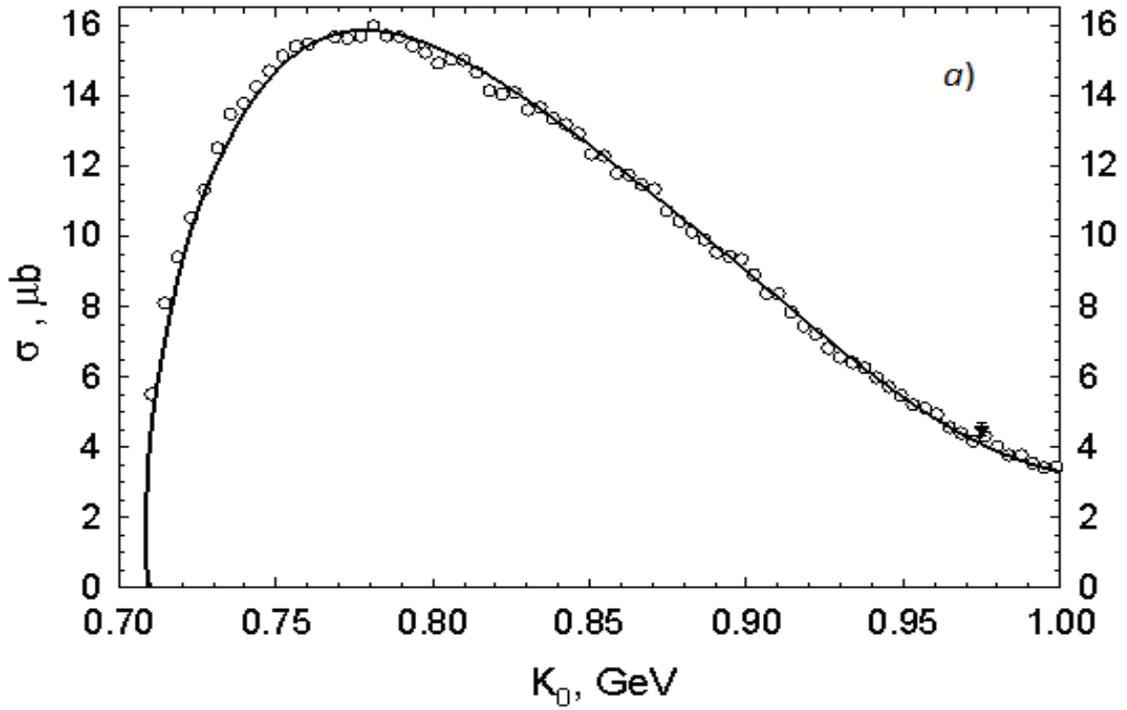

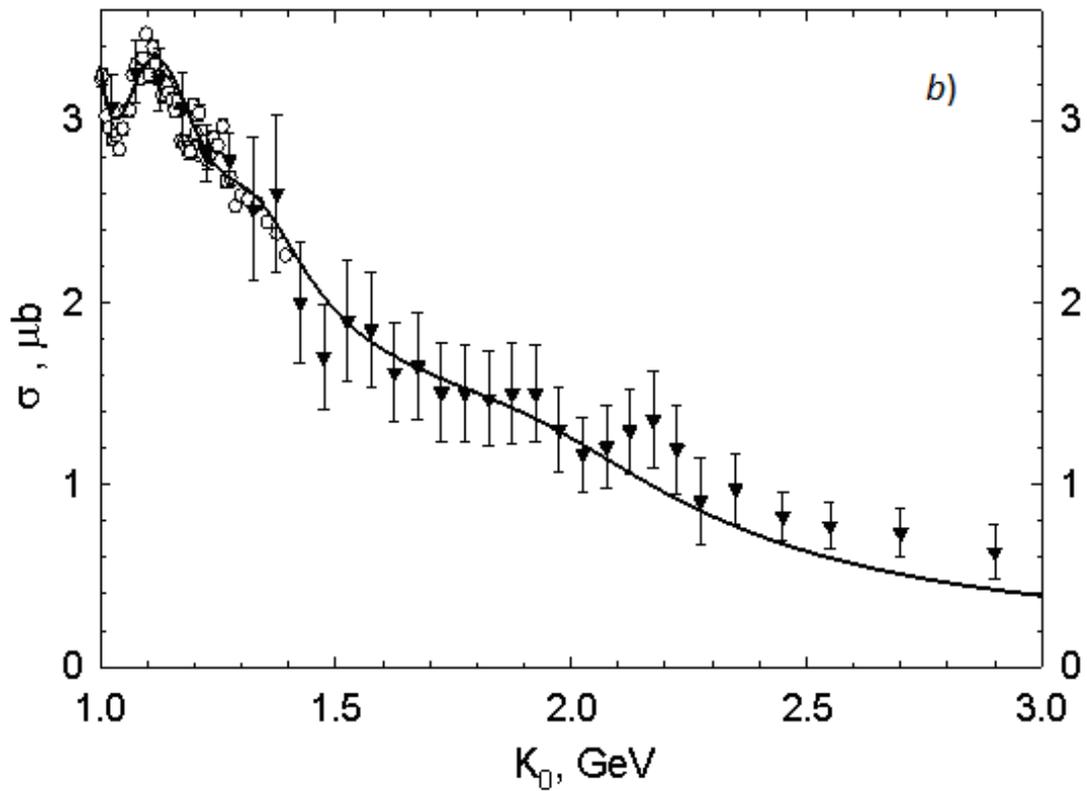

Fig. 3. Total cross section for the reaction $\gamma p \to \eta p$. The data are from Ref. [8] (○) and [3] (▼).

Figs. 3a and 3b show the total cross section of the reaction (1) which is compared with the data taken only from [3] and [8]. The total cross sections, obtained from the measured angular distributions of the earlier works are in rather good agreement with those given in [3] and [8]. In Fig. 4 the total cross section for $\gamma p \to \eta p$ is shown as function of the photon laboratory energy. The different curves are obtained after one of the contributing resonances listed in Table 1 is excluded from the amplitude. One readily notes a crucial role of the resonances $S_{11}(1535)$, $S_{11}(1650)$, and $P_{13}(1720)$.

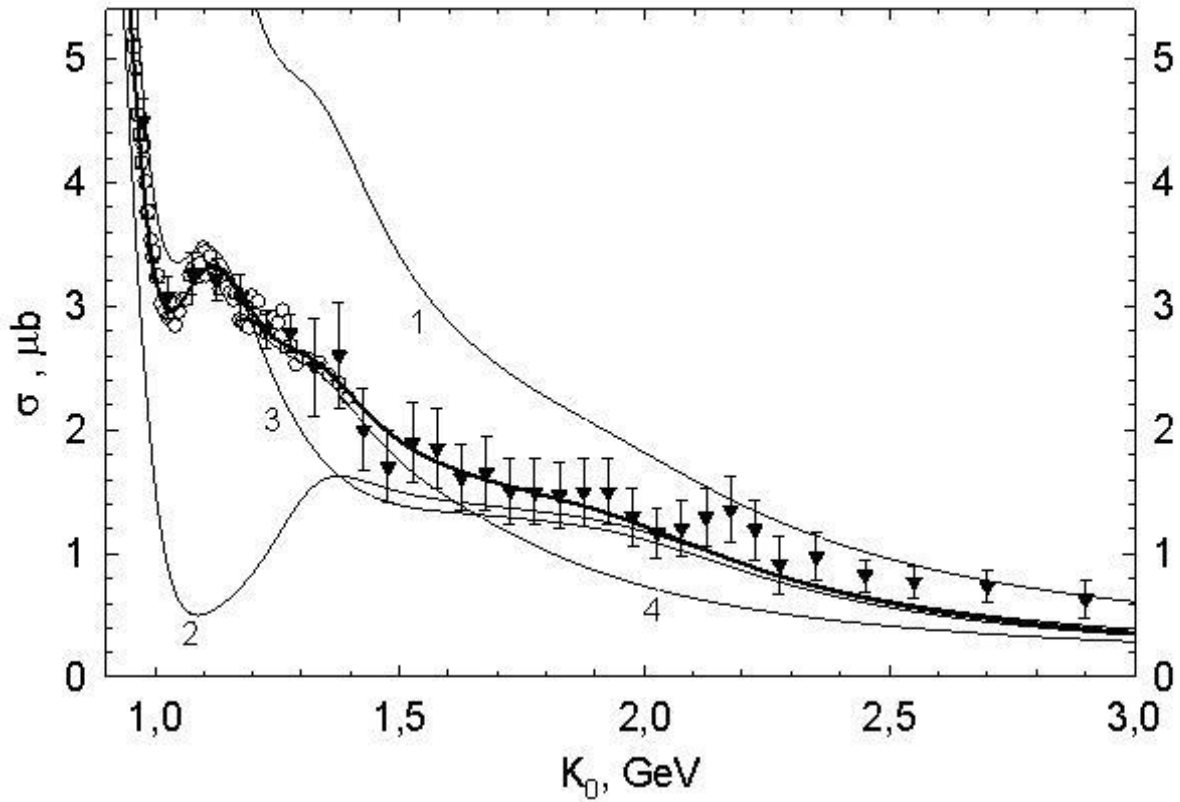

Fig. 4. Total cross section for $\gamma p \to \eta p$. In the thick solide curve the contribution of all resonances from Table 1 are included. In each thin solide curve the contribution of one resonance in Table 1 was excluded: $1 - S_{11}(1650)$; $2 - P_{13}(1720)$; $3 - S_{11}(1830)$; $4 - D_{15}(2150)$. The data are from Ref. [8] (○) and [3] (▼)

In Fig. 5 we also show $T$- and $F$-asymmetry for $\gamma p \to \eta p$ and compare them with the preliminary data from Ref. [9] in the region up to the lab photon energy 1 GeV. As may be seen, whereas in the lower energy region ($K_0 < 1000$ MeV) there is at least qualitative agreement between the theoretical and the experimental results, at higher energies the calculated $T$- and $F$-asymmetries [30] disagree with the data [9].

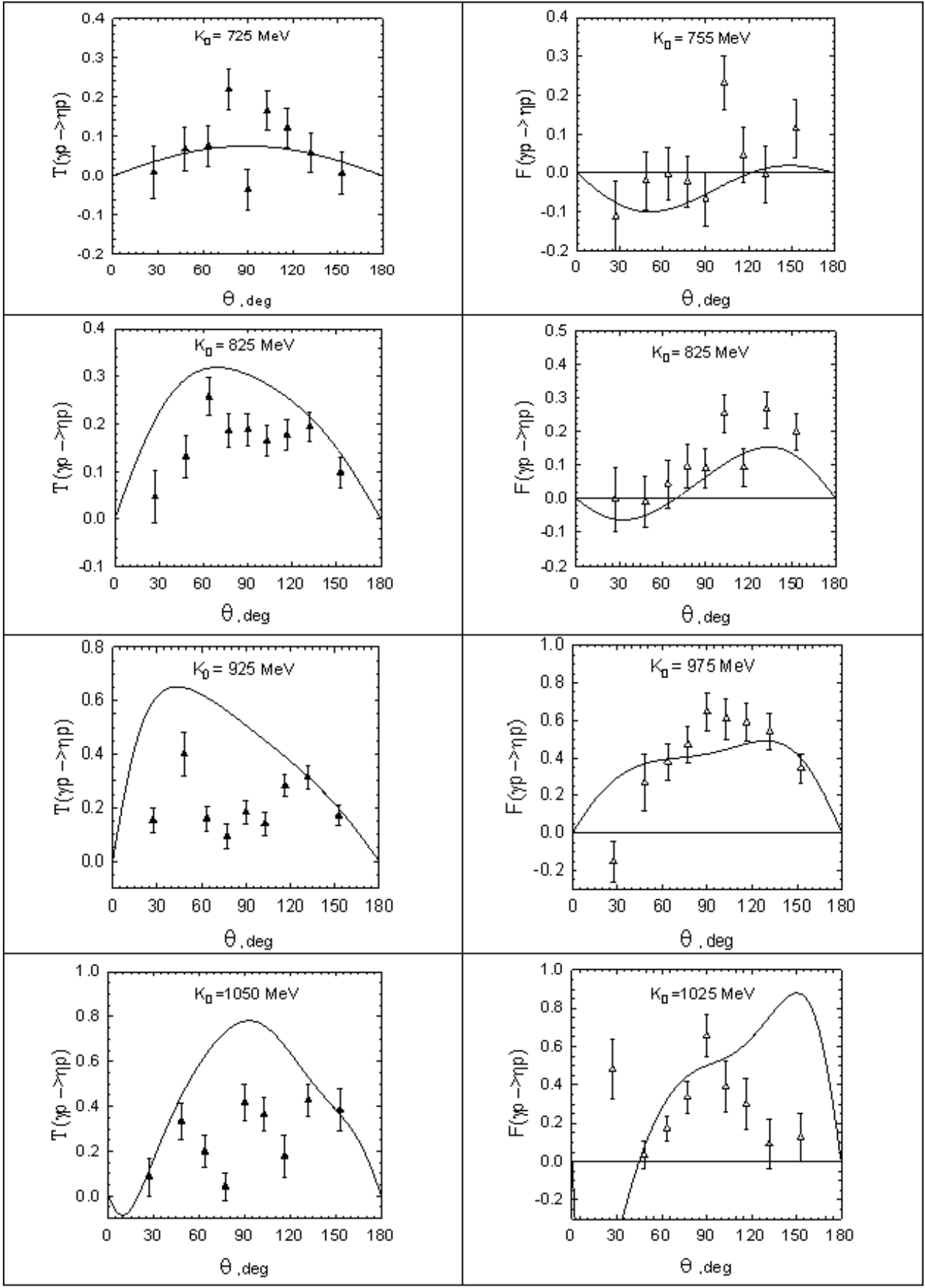

Fig. 5. *T* and *F* asymmetries for $\gamma p \to \eta p$. The data are from Ref. [9]

Comprehensive information about the properties of nucleon resonances may be obtained, when the photo- and electroproduction processes both on protons and neutrons are investigated. Use of the deuteron targets together with coincident measurements allow one to study the process $\gamma n \to \eta n$ using quasifree photoproduction of η-mesons on the bound neutrons. In this way the differential cross section [10, 11] as well as Σ-asymmetry [12] for the process (2) were obtained. Furthermore, during the preparation of the present paper new data were obtained at Mainz Microtron [31, 32], which are not included into our analysis.

To describe the process $\gamma n \to \eta n$ in the energy region considered, we included the resonances, whose masses $W_r$ and widths $\Gamma_r$ were obtained in [25, 26] for $\gamma p \to \eta p$ (see Table 1). Using these values we adjusted the electromagnetic couplings $\gamma^E$ and $\gamma^M$ in such a way that good description of the experimental unpolarized differential cross section [11] as well as Σ-asymmetry [12] for $\gamma n \to \eta n$ is achieved. The overall sign of the $P_{11}(1440)$ term turned out to be opposite to that in the channel $\gamma p \to \eta p$.

**Table 2**. Electromagnetic parameters of the resonances listed in Table 1, obtained in the present work by fitting the experimental data of the process $\gamma n \to \eta n$.

| $N^*$-resonances | $W_r$, MeV | $\Gamma_r$, MeV | $\gamma^E$, MeV | $\gamma^M$, MeV | $\xi_{1/2}$, $10^{-1}$ GeV$^{-1}$ | $\xi_{3/2}$, $10^{-1}$ GeV$^{-1}$ |
|---|---|---|---|---|---|---|
| $S_{11}(1535)$ | 1532 | 161 | -1.556 | — | -1.741 | — |
| $S_{11}(1650)$ | 1642 | 140 | -0.068 | — | -0.087 | — |
| $S_{11}(1830)$ | 1828 | 155 | -0.060 | — | -0.069 | — |
| $P_{11}(1440)$ | 1440 | 350 | — | -0.250 | — | — |
| $P_{13}(1720)$ | 1724 | 173 | 0.050 | -0.500 | -0.262 | -0.580 |
| $D_{13}(1520)$ | 1520 | 120 | -0.400 | -0.150 | -0.038 | -0.151 |
| $D_{15}(2150)$ | 2140 | 382 | 0.300 | -0.500 | -0.358 | 0.111 |
| $F_{15}(1680)$ | 1685 | 130 | -0.200 | -0.050 | -0.064 | -0.132 |
| $F_{17}(1990)$ | 1999 | 312 | 0.250 | -0.425 | -0.039 | -0.306 |

The results of our fit are presented in Table 2 and in Figs. 6 to 10. In Fig. 6 the calculated differential cross sections are compared with those measured in Ref. [11]. It is worth noting that rather good description at the energies $K_0 > 1600$ MeV is achieved only through inclusion of the resonances $F_{17}(1990)$ and $D_{15}(2150)$, which were previously introduced into the amplitude of the process (1). This result may be viewed as an indication of an existence of these states in the nucleon spectrum.

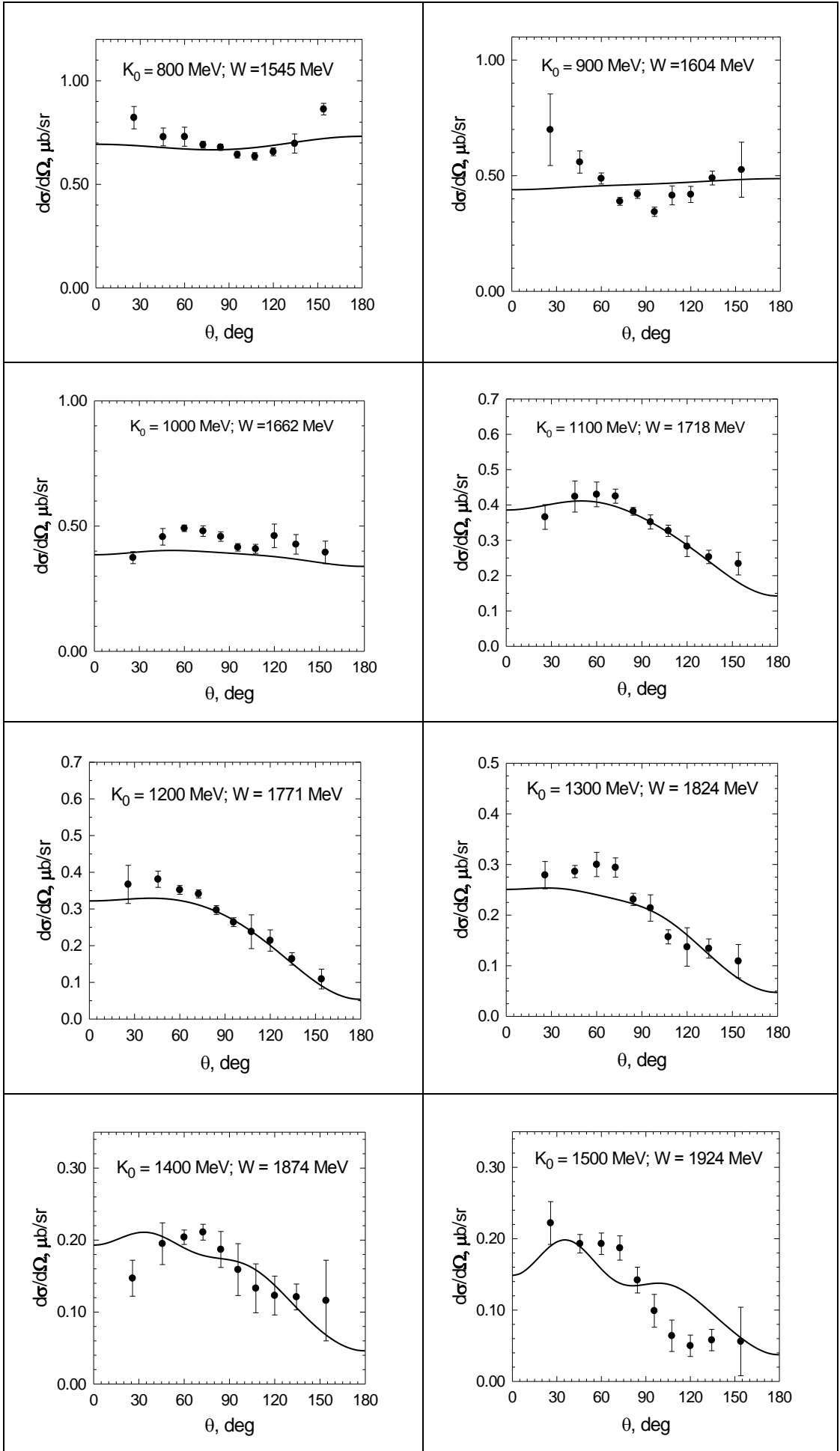

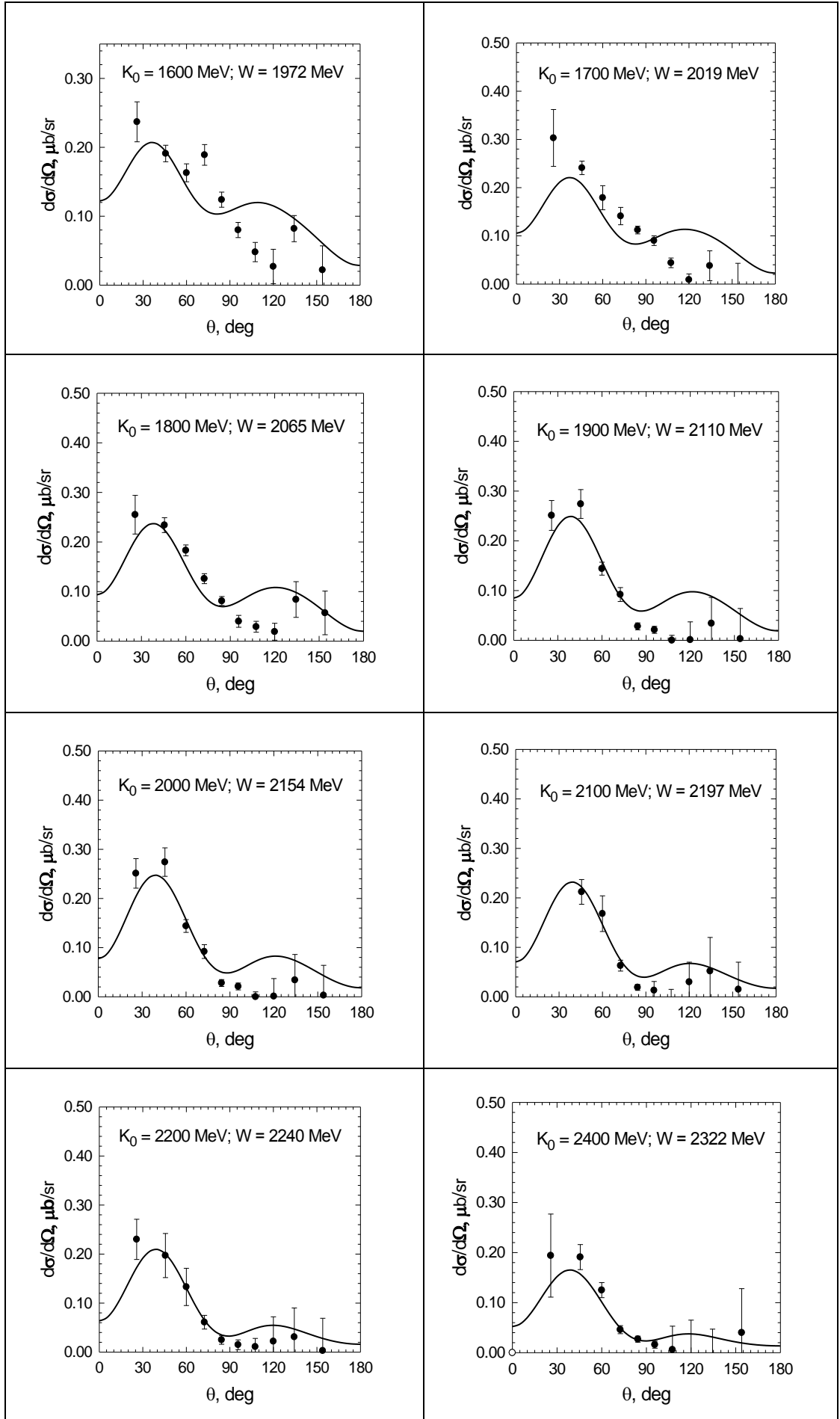

Fig. 6. Same as in Fig.1, but for $\gamma n \to \eta n$. The data are from Ref. [11]

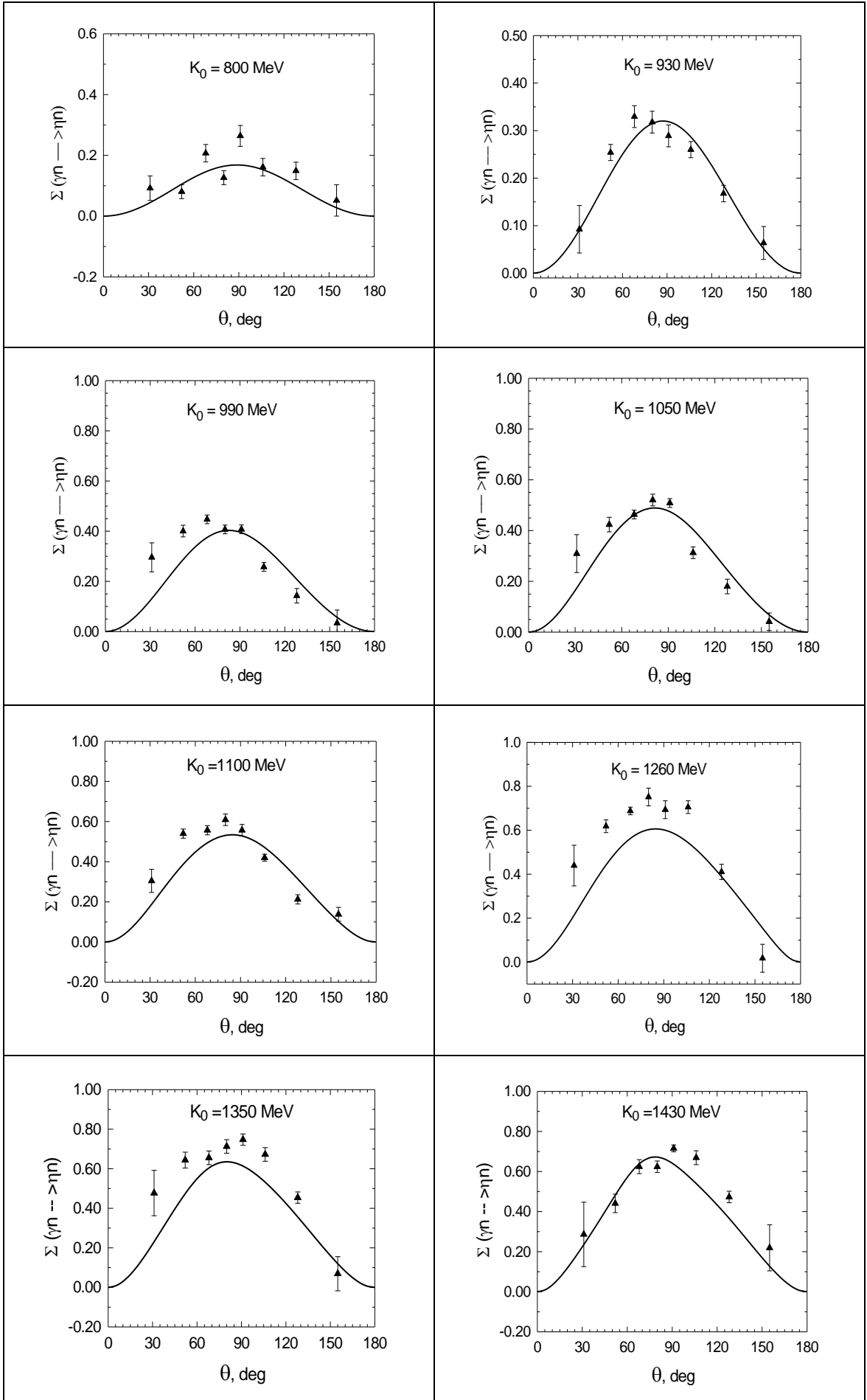

Fig. 7. $\Sigma$–asymmetry of the reaction $\gamma n \to \eta n$. The data are taken from Ref. [12]

In Fig. 7 we demonstrate angular distributions of Σ-asymmetry for the process (2) for different energies of the incident photons. It is worth to note, that this observable is very similar for η photoproduction on protons and neutrons: in both cases it is positive and rather large in magnitude in the energy region considered (see Figs. 2 and 7).

As one can see from Figs. 6 and 7, our calculation of the differential cross section and Σ-asymmetry for $\gamma n \to \eta n$ is in reasonable agreement with the data. The agreement is slightly better in comparison with the results of the theoretical works [33, 36]. Furthermore, as was already noted above, in our fit we did not use the resonances $P_{11}(1710)$ and $D_{15}(1675)$ which are sometimes found in the processes (1) and (2) [28, 33, 35, 36].

Originally, one of the main ideas of the present paper was to show that the processes (1) and (2) can be described only with three resonances

$$S_{11}(1535), \quad S_{11}(1650) \text{ and } P_{13}(1720).$$

However, as may be seen from the last two columns in Table 2, the cross sections on the neutron are saturated already with two resonances

$$S_{11}(1535), P_{13}(1720)$$

in the corresponding energy region. The question about the role of $S_{11}(1650)$ to the reaction (2) seems to be not as trivial as was initially expected.

As has long been known, the $S_{11}(1535)$ resonance is mostly excited by the isovector photons [11, 33, 34] and, as a consequence, it enters the amplitudes of the processes (1) and (2) with opposite signs (see Tables 1 and 2). Furthermore, it was established that the contributions of $S_{11}(1535)$ and $S_{11}(1650)$ interfere destructively in the amplitude (1) [11, 33, 35, 36, 37]. This interference manifests itself as a visible minimum in the $\gamma p \to \eta p$ total cross section (see Fig. 4). If the resonance $S_{11}(1650)$ has predominantly isoscalar nature, then these two s-wave resonances should interfere constructively in the amplitude (2). This in turn should lead to a second maximum (with very weak evidence of a minimum) in the $\gamma n \to \eta n$ total cross section [11] at the energy around $W \approx 1660$ MeV (Fig. 8). At the same time, appearance of the second maximum in the total cross section (2) implies existence of a minimum, so that these resonances may interfere destructively also in the process (2) because of opposite signs of their contributions [28]. This point is interesting since it may indicate that the resonance $S_{11}(1650)$ also has an isovector nature as do the nearly lying resonance $S_{11}(1535)$.

The contributions of $D_{13}(1520)$ and $F_{15}(1680)$ influence only weakly the total cross sections of both reactions, but they appear to be rather visible in angular distributions of the produced mesons as well as in the polarization observables. We would also like to note that we were able to describe η photoproduction without using such well-established resonances as $P_{11}(1710)$ and $D_{15}(1675)$, which according to the results of several works [28, 33, 35, 36] may contribute to the processes (1) and (2). Perhaps, the pairs of resonances $P_{11}(1710)$, $D_{15}(1675)$ and $P_{13}(1720)$, $F_{15}(1680)$ simply duplicate each other in $\gamma N \to \eta N$ and further investigation should include analysis of polarization observables which are more sensitive to the partial wave content of the amplitude.

In Fig. 8 we present total cross section of the process (2) as function of the incident photon energy.

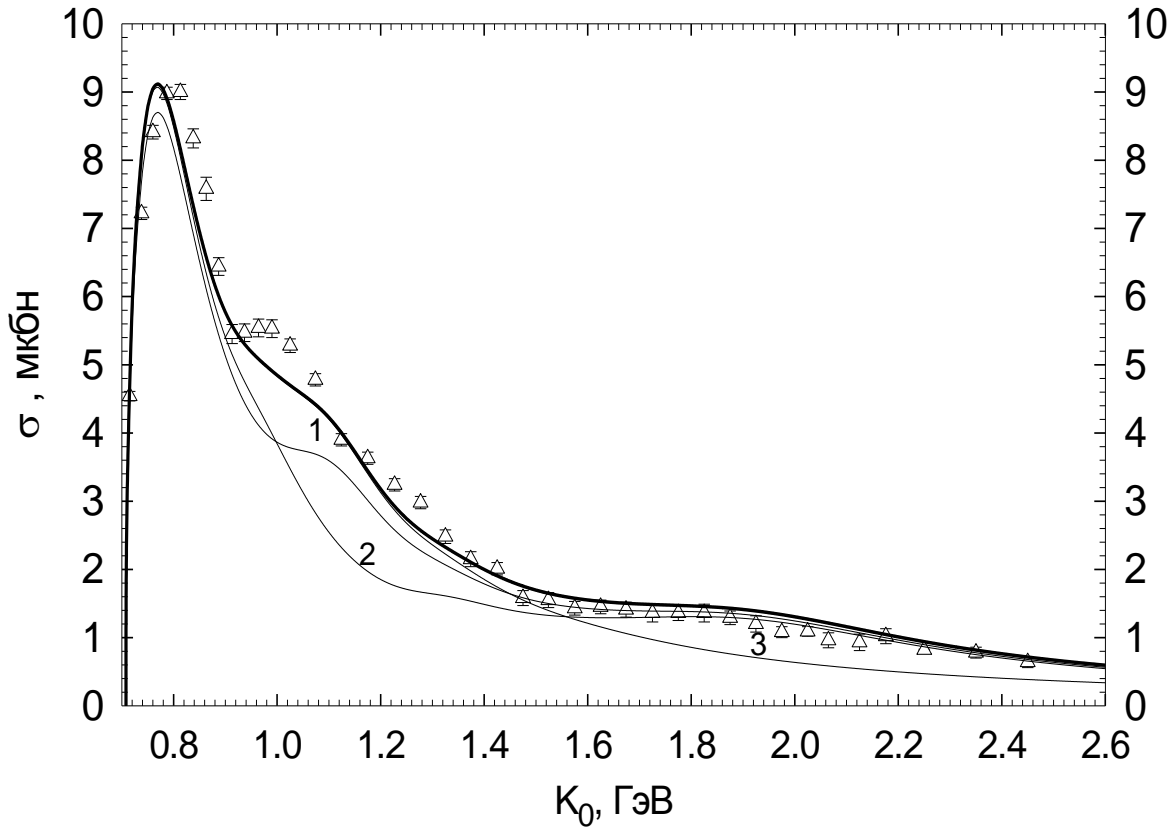

Fig. 8. Total cross section of the process $\gamma n \to \eta n$. The thick solid curve contains all resonance s from Table 2. The curves 1, 2 and 3 are obtaind when one of the resonences in Table 2 is excluded: 1 − $S_{11}(1650)$; 2 − $P_{13}(1720)$; 3 − $D_{15}(2150)$; Δ − experimental data from Ref. [11]

In Fig. 8 we also present our results for the total cross section of the process $\gamma n \to \eta n$, which were obtained after elimination of individual resonances with parameters from Table 2. The results are presented for successive elimination of only three resonances, since exclusion of other resonances ($S_{11}(1830)$, $D_{13}(1520)$, $F_{15}(1680)$, and $F_{17}(1990)$) does not visibly change the resulting cross section. As one can see, there is a constructive interference of $S_{11}(1535)$ and $S_{11}(1650)$ and there is visible contribution of the resonance $P_{13}(1720)$.

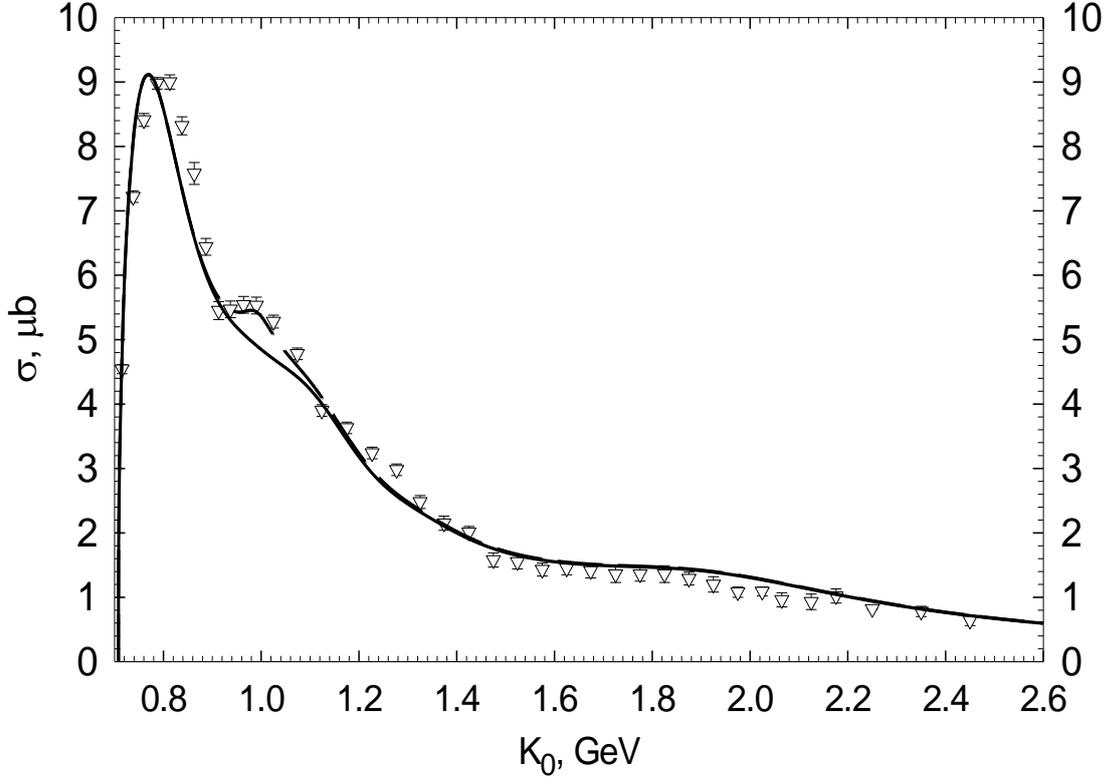

Fig. 9. Total cross sections for the reaction $\gamma n \to \eta n$ calculated with parameters listed in Table 2 (solid curve) and with narrow resonance $P_{11}(1670)$ (dashed curve); $\nabla$ − experimental data from Ref. [11]

In Fig. 9 we present total cross section of the process (2) as function of the incident photon energy. The second slight maximum at the energy close to 1 GeV is not well reproduced if only constructive interference of $S_{11}(1535)$ and $S_{11}(1650)$ and other resonances from Table 2 is taken into account (Fig.8 thick curve). However the latest experimental results have confirmed that rather narrow maximum in this region appears also in η photoproduction on the light nuclei [31, 32]. Therefore, we included into the amplitude the exotic state $P_{11}(1670)$ which seems to be excited only on neutrons [10, 11, 31, 32] (Fig. 9 dashed curve). Absence of the other $P_{11}$ resonances in our model allows one to identify the contribution of the exotic resonance in the $\gamma n \to \eta n$ total cross section and in this way to determine its properties rather precisely. In particular we find

$$W_r = 1656 \text{ MeV}, \ \Gamma_r = 51 \text{ MeV}, \gamma^M = -0.119 \text{ MeV}. \qquad (4)$$

However inclusion of the exotic state $P_{11}(1670)$ deteriorates the quality of our previous fit. In particular, moderate quality of our description of the data at $K_0 = 1000$ MeV with resonances listed in Table 2 (solid curve in Fig. 10) becomes worse after inclusion of the narrow $P_{11}$ state (dashed curve in the same figure).

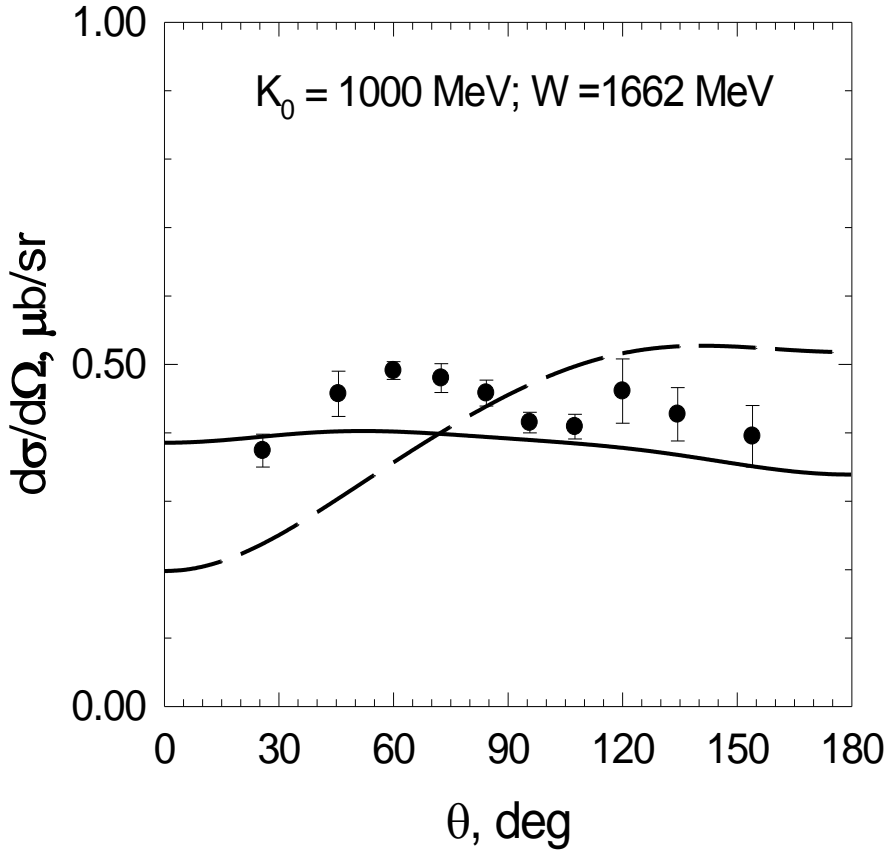

Fig. 10. Differential cross section of the reaction $\gamma n \to \eta n$ with resonances from Table 2 (solid curve) and with addition of the narrow resonance $P_{11}(1670)$ + resonances from Table 2 (dashed curve), ● is experimental results from Ref. [11]

Perhaps, this structure in the total cross section for $\gamma n \to \eta n$ may be described better in the *s*-wave, as has been done in Ref. [31]. However, in this case we will face with undesirable interference of four *s*-wave resonances.

The quantities $\xi_\lambda$ are directly related to the helicity amplitudes $A_\lambda$ as is given in (3). This allows one to estimate the amplitude $A_{1/2}$ of exotic state $P_{11}(1670)$:

$$\xi_{1/2} = -0.244 \cdot 10^{-1} \text{ GeV}^{-1}; \quad A^n_{1/2} = -18.4 \cdot 10^{-3} \text{ GeV}^{-1/2}, \qquad (5)$$

if the ratio of the partial $\eta n$ decay width to the total width is taken as 0.1, and

$$A^n_{1/2} = -5.80 \cdot 10^{-3} \text{ GeV}^{-1/2}, \qquad (5)*$$

if this ratio is equal to 1.

For the ratio of the helicity amplitudes on protons and neutrons one obtains from Tables 1 and 2

$$S_{11}(1535): \qquad A^n_{1/2}/A^p_{1/2} = -0.691. \qquad (6)$$

$$S_{11}(1650): \qquad A^n_{1/2}/A^p_{1/2} = +0.089. \qquad (7)$$

$$S_{11}(1830): \qquad A^n_{1/2}/A^p_{1/2} = -0.314. \qquad (8)$$

The ratio (6) is in reasonable agreement with the existing experimental and theoretical results. Since the information about the value of $A^n_{1/2}$ for the resonance $S_{11}(1535)$ [28, 33] is rather contradictory, we give here the value coming out of our fit

$$A^n_{1/2} = -76 \cdot 10^{-3} \text{ GeV}^{-1/2}, \qquad (9)$$

which is very close to the value found in Ref. [33] and is in agreement with the results of earlier multipole analyses of pion photoproduction on neutrons [38]. However, the absolute value of $A^n_{1/2}$ given by (9) is smaller than that obtained in Ref. [28].

The ratio $A^n_{1/2}/A^p_{1/2}$ for the resonance $S_{11}(1650)$ (7) is positive. According to our model, this result indicates that this resonance is mostly excited by the isoscalar photons. The positive value of the ratio $A^n_{1/2}/A^p_{1/2}$ for $S_{11}(1650)$ was also obtained in the energy independent multipole analysis for the processes $\gamma N \to \pi N$ in Ref. [38], however as was already mentioned above, in Ref. [28] and in the compilation [29] this value is negative. The contribution of the resonance $S_{11}(1830)$ [39] into the process (2) is significantly smaller than its contribution to the process (1) (see Tables 1 and 2). Furthermore,

$$P_{13}(1720): \qquad A^n_{1/2}/A^p_{1/2} = -0.981 \qquad A^n_{3/2}/A^p_{3/2} = -0.872 \qquad (10)$$

$$D_{13}(1520): \qquad A^n_{1/2}/A^p_{1/2} = 0.594 \qquad A^n_{3/2}/A^p_{3/2} = -1.269 \qquad (11)$$

$$D_{15}(2150): \qquad A^n_{1/2}/A^p_{1/2} = -1.874 \qquad A^n_{3/2}/A^p_{3/2} = -0.776 \qquad (12)$$

$$F_{15}(1680): \qquad A^n_{1/2}/A^p_{1/2} = -1.778 \qquad A^n_{3/2}/A^p_{3/2} = -1.023 \qquad (13)$$

$$F_{17}(1990): \qquad A^n_{1/2}/A^p_{1/2} = -0.280 \qquad A^n_{3/2}/A^p_{3/2} = -1.987 \qquad (14)$$

Our results for the ratios (10) is in general agreement with *MAID*, confirming that the resonance $P_{13}(1720)$ is mostly excited by the isovector photons. At the same time, this result contradicts to the data recommended by [29]. The calculated ratios (11), (13) and (14) are in qualitative agreement with the compilation [29].

Quantitative disagreement of the ratios (7), (10–11, 13–14) with those from [29] is rather peculiar for many analyses of meson photoproduction on protons and neutrons and probably indicates that our knowledge about their electromagnetic properties is still incomplete. We also would like to note that in the latest Particle-Data-Group compilation [40] these ratios are not presented.

In summary, using the isobar model for the process $\gamma p \to \eta p$ whose parameters were fitted to the existing data we obtained rather good description of $\eta$ photoproduction on neutrons without the latest data for photoproduction on the bound nucleons [31, 32]. As a result, we performed energy dependent multipole analysis of $\eta$ photoproduction on nucleons.

Our main results may be summarized as follows: (i) the main contribution to the cross section of the process $\gamma p \to \eta p$ comes from the resonances $S_{11}(1535)$, $S_{11}(1650)$, $P_{13}(1720)$, and only from two of them to the cross section for $\gamma n \to \eta n$; (ii) the resonances $D_{13}(1520)$ and $F_{15}(1680)$ are needed to reproduce the shape of the measured angular distributions, as well as to describe $\Sigma$-asymmetry in both charge channels (1) and (2), however their contribution to the total cross section is insignificant; (iii) the contribution of the Roper resonance $P_{11}(1440)$ has different sign in the proton and the neutron channels; (iv) our analysis gives predictions for the ratios of the helicity amplitudes for photoexcitation of the resonances on protons and neutrons; (v) at higher energies the form and the magnitude of the differential cross sections are mostly determined by the resonance $D_{15}(2150)$; (vi) there might exist the exotic resonance $P_{11}(1670)$, which would explain the second narrow maximum in the $\gamma n \to \eta n$ total cross section. Otherwise, in order to explain the appearance of this maximum, one has to investigate $\eta$ photoproduction on nuclei [31, 32]; (vii) we did not need to include the nonresonant background, as was discussed above in the main part of the paper.

I would like to thank A. Fix for assisting in the preparation of this article.